\documentclass[epsf]{aa}
\usepackage{natbib}
\usepackage{graphicx}

\begin{document}

\title{CK Vul as a Candidate Eruptive Stellar Merging Event}
\subtitle{}
\authorrunning{T. Kato}
\titlerunning{CK Vul as a Candidate Eruptive Stellar Merging Event}

\author{Taichi Kato\inst{1}}

\institute{
  Department of Astronomy, Kyoto University, Kyoto 606-8502, Japan
}

\offprints{Taichi Kato, \\ e-mail: tkato@kusastro.kyoto-u.ac.jp}

\date{Received / accepted }

\abstract{
   CK Vul (Nova Vul 1670) is one of the most mysterious objects among
erupting stellar objects.  Past studies have suggested that CK Vul
is a final helium-flash object resembling V605 Aql and V4334 Sgr
(Sakurai's object).  The peculiar outburst light curve of CK Vul, however,
had no similar counterpart among the known eruptive objects.  Furthermore,
the presence of hydrogen in the proposed remnant seems to contradict
with the final helium-flash scenario.  We propose that the peculiarities
of CK Vul can be naturally understood if we consider a merging of
main-sequence stars, following a new interpretation by \citet{sok02v838mon}
which was proposed to explain the peculiar eruptive object V838 Mon.
In this case, the 1670 outburst of CK Vul may be best understood as
a V838 Mon-like event which occurred in our vicinity.
}

\maketitle

\section{Introduction}

   CK Vul (Nova Vul 1670) is one of the most mysterious objects among
erupting stellar objects.  The object reached a maximum of $m_v$ = 2.6,
with at least three brightness peaks spanning for three years.
The overall light curve and the multiple brightness peaks spanning for
such a long period is unlike any classical nova
\citep{GalacticNovae,due81novatype,due87novaatlas}.  The object was
claimed to be recovered as the ``oldest old nova'' \citep{sha82ckvul}
from the presence of a planetary nebula-like shell.  \citet{sha85ckvul}
further proposed, from the low absolute magnitude of the supposed
quiescent counterpart, that CK Vul is a ``hibernating'' old nova
(cf. \cite{Hibernation}).

   \citet{har96ckvulv605aql}, however, performed near-infrared photometry
of the suggested counterpart, and yielded a spectral energy distribution
(SED) indicating the presence of a dust shell.  From the similarity of
SED and the outburst light curve with those of V4334 Sgr (Sakurai's object),
\citet{har96ckvulv605aql} suggested that CK Vul may be related to V4334 Sgr,
which is believed to be a final helium-flash object
\citep{due97v4334sgr,ker99v4334sgr} rather than a classical nova.
\citet{eva02ckvul} presented further evidence for a final helium-flash
object from the detection of a cavity in IRAS infrared map and the analysis
of sub-mm emission.  The identification of the true remnant of CK Vul
seems to be established by this study.

   However, it is controversial whether the unique light curve of CK Vul
can be reasonably explained by the scenario of a final helium-flash
object.  Although both V605 Aql and V4334 Sgr had multiple optical peaks, the
later peaks were always fainter than the preceding peaks.  This phenomenon
is natural since the optical variations are regarded as a consequence
of a combination of continuous cooling and dust formation
\citep{asp99v4334sgr}.  As pointed out by \citet{sha85ckvul}, a nova-type
thermonuclear runaway event should decay with a dynamical time-scale
($\sim$100 d) from a super-Eddington maximum phase, which further makes
it difficult to explain the long intervals between the brightenings.

\begin{figure*}
  \begin{center}
  \includegraphics[angle=0,height=12cm]{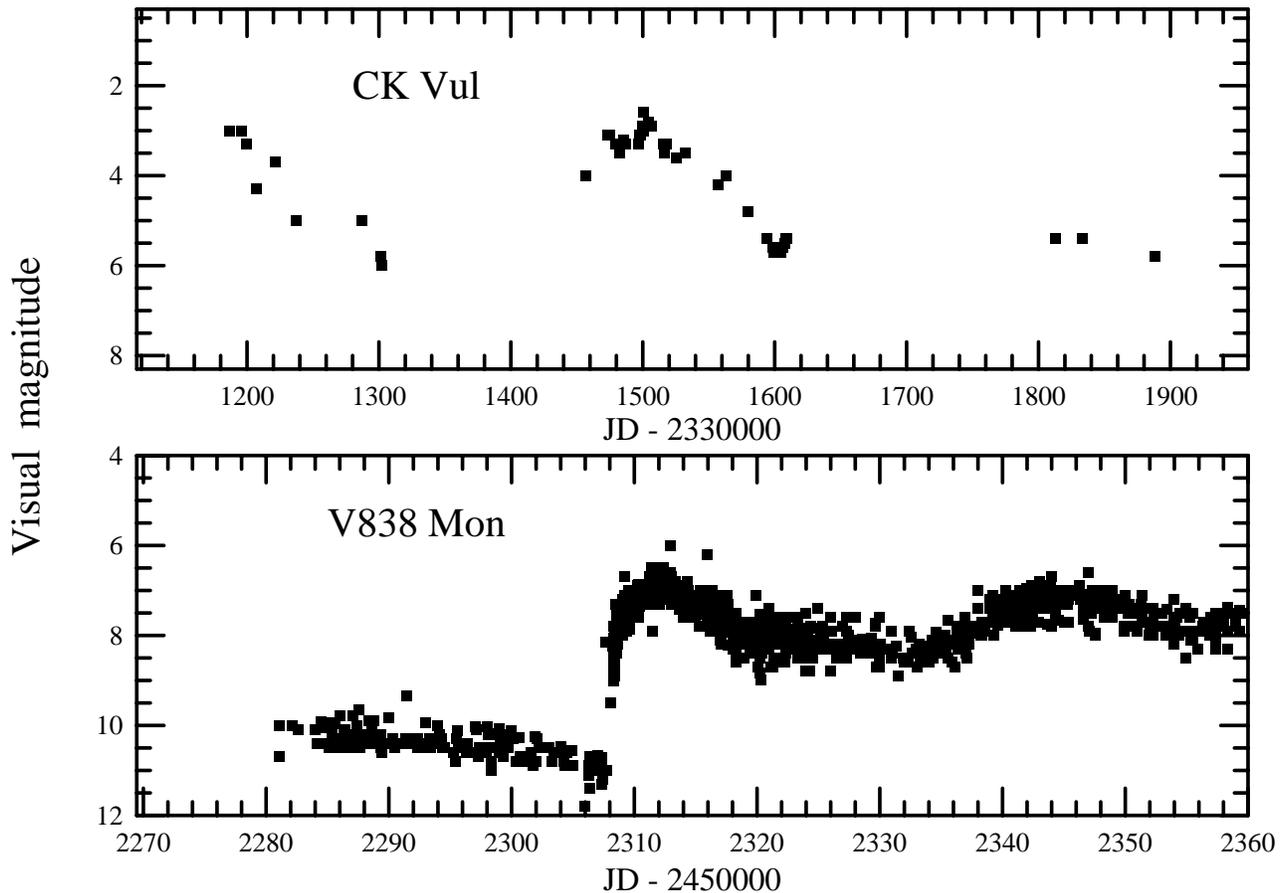}
  \end{center}
  \caption{Comparison of the light curves of CK Vul and V838 Mon.
  The data of CK Vul are from \citet{sha85ckvul}.  The data of V838 Mon
  (visual and $V$-band observations) are from the reports to VSNET.}
  \label{fig:comp}
\end{figure*}

   Most recently, a unique eruptive object, V838 Mon, has been discovered
\citep{bro02v838moniauc7785}.  The subsequent evolution of V838 Mon,
with multiple optical peaks and very cool atmosphere \citep{mun02v838mon}
has been remarkably unlike any class of erupting
stellar objects.  The light curve of V838 Mon in part resembles that of
CK Vul, particularly in its second bright maximum long time after the
initial eruption (Fig. \ref{fig:comp}).
Although the exact mechanism of the eruption of V838 Mon
has not been clarified, \citet{sok02v838mon} proposed that a merging of
two main-sequence stars can explain the unique outburst of V838 Mon.
Based on the phenomenological resemblance between CK Vul and V838 Mon,
we try to explore the possibility of explaining CK Vul within the
framework of \citet{sok02v838mon}.

\section{Application to CK Vul}

   According to \citet{sok02v838mon}, the eruption of V838 Mon has been
proposed to be a merging of a close main-sequence binary with $M_1 \sim$
1.5 $M_\odot$ and $M_2 \sim$ 0.1--0.5 $M_\odot$.  The first maximum
was explained as a runaway mass-transfer when the primary (more massive)
star fills the Roche-lobe.  The observed luminosity of the initial maximum
can be well reproduced by a reasonable parameter in a calculation of
an accreting 0.2 $M_\odot$ main-sequence star \citep{pri85MSaccretion}
and the subsequent expansion to a giant.  \citet{sok02v838mon} proposed
that the second (brighter) maximum can be explained by the accretion of
the secondary onto the primary, particularly by the merging of the stellar
cores, which enables a further expansion \citep{hje91MSaccretion}.

   The scenario by \citet{sok02v838mon} thus seems to able to explain
the reason why the second maximum of V838 Mon was more energetic than the
first maximum.  In CK Vul \citep{sha85ckvul}, the second maximum was recorded
to be at least as bright as the first (initially discovered) maximum.
In \citet{sok02v838mon}, the brightness of the first maximum depends on
the parameter $\alpha$, which is a fraction of the free-fall energy
employed to expand the atmosphere.  In reproducing V838 Mon,
\citet{sok02v838mon} used $\alpha$ = 0.1, while a higher value (up to
$\alpha$ = 0.5) is realistic.  The time-scales between the first and
second maximum can be strongly different depending on the initial condition
of the binary and the degree of angular momentum loss, both of which are
not still reasonably constrained.

   A subtle difference between the light curves of CK Vul and V838 Mon
is not thus problematic within the scenario by \citet{sok02v838mon}.
A more remarkable consequence is that this scenario will be able to explain
the hydrogen-rich appearance \citep{nay92Hibernation} of the proposed
remnant object of CK Vul, which is unlike a hydrogen-poor Wolf-Rayet-like
object \citep{cla97v605aql}, R CrB-like object, or cooling white dwarf
what are expected from the final helium-flash scenario.

   The nebula surrounding CK Vul
\citep{sha82ckvul,sha85ckvul} is also unusual for a planetary nebula.
\citet{sha85ckvul} suggested, from an analysis of emission lines and
an estimate of the recombination time, that the nebula is excited by
a shock which was produced by a collision of the CK Vul wind (or ejecta)
with the interstellar matter.  Such a feature has not been observed
in final helium-flash objects, while an expected heavy mass-loss
from a merging binary can more reasonably explain the origin.
An estimated nebular expansion velocity of 59$\pm$16 km s$^{-1}$ is
reasonable compared to the outflows with the observed velocities of
200--500 km s$^{-1}$ in V838 Mon considering an effect of deceleration.
Such an outflow was not observed in the final helium-flash object,
V4334 Sgr \citep{due97v4334sgr}.

   Finally, we compare the outburst absolute magnitudes between CK Vul
and V838 Mon.  With a long-distance estimate ($\sim$3 kpc, H. E. Bond,
cited in \cite{sok02v838mon}),\footnote{
  The estimated distance of V838 Mon significantly differs between the
  authors.  For example, \citet{mun02v838mon} suggested a distance of
  790$\pm$30 pc based on an assumption of the circumstellar origin of
  the light echo.
  The true distance is, however, still controversial (e.g.
  \cite{mun02v838moniauc8005} even discussed a possibility of 10.5 kpc).
  Since the variety of parameters in \citet{sok02v838mon}
  can naturally reproduce a diverse range of energetics, the ambiguity
  of the distance of V838 Mon would not be a serious problem in interpreting
  CK Vul itself, although the deduced similarity with V838 Mon might
  become less striking.
}
and a reasonable extinction of $A_V \sim$
1.5 \citep{mun02v838mon}, the maximum $M_V$ of V838 Mon is estimated
to be $-$7.0.  This value is appreciably close to $M_V \sim -8$ by
adopting the estimated distance of 550 pc and of $A_V \sim$ 2.2 in CK Vul
\citep{sha85ckvul}.  \citet{sha85ckvul} noted that some of Helvelius'
1670--1671 observations described CK Vul as `blurred'.  This intriguing
finding may be somehow related to a V838 Mon-like light echo
\citep{mun02v838mon}, although it seems rather unlikely to resolve such
a light echo with the unaided eyes even at a small distance.
Considering the ambiguity of the distance
estimates of these objects, the 1670 outburst of CK Vul may be best
understood as a V838 Mon-like event which occurred in our vicinity.
Future detailed observations of the CK Vul remnant in view of the
present possibility could provide a crucial test for this interpretation.

\vskip 1.5mm

We are grateful to the referee (A. Evans) for drawing the author's
attention to the description of Helvelius' 1670--1671
observations of CK Vul.
We are grateful to the observers who reported vital observations
to VSNET ({\tt http://www.kusastro.kyoto-u.ac.jp/vsnet/}).
This work is partly supported by a grant-in-aid (13640239)
from the Japanese Ministry of Education, Culture, Sports,
Science and Technology.

\end{document}